\documentclass{article}

\usepackage{arxiv}

\usepackage[utf8]{inputenc} % allow utf-8 input
\usepackage[T1]{fontenc}    % use 8-bit T1 fonts
\usepackage{hyperref}       % hyperlinks
\usepackage{url}            % simple URL typesetting
\usepackage{booktabs}       % professional-quality tables
\usepackage{amsfonts}       % blackboard math symbols
\usepackage{nicefrac}       % compact symbols for 1/2, etc.
\usepackage{microtype}      % microtypography
\usepackage{multirow}
\usepackage{lipsum}
\usepackage{graphicx}
\usepackage{amsmath}
\usepackage{nccmath}
\usepackage[super]{nth}
\graphicspath{ {./images/} }

\title{Quantifying Community Resilience Based on Fluctuations in Visits to Point-of-Interest from Digital Trace Data}

\author{
 Cristian Podesta \\
  Urban Resilience.AI Lab\\
  Zachry Department of Civil and \\Environmental Engineering\\
  Texas A\&M University\\
  College Station, TX 77843 \\
  \texttt{cristianapodesta@tamu.edu} \\
   \And
 Natalie Coleman \\
  Urban Resilience.AI Lab\\
  Zachry Department of Civil and \\Environmental Engineering\\
  Texas A\&M University\\
  College Station, TX 77843 \\
  \texttt{ncoleman@tamu.edu} \\
  \And
 Amir Esmalian \\
  Urban Resilience.AI Lab\\
  Zachry Department of Civil and \\Environmental Engineering\\
  Texas A\&M University\\
  College Station, TX 77843 \\
  \texttt{amiresmalian@tamu.edu} \\
  \And
 Faxi Yuan \\
  Urban Resilience.AI Lab\\
  Zachry Department of Civil and \\Environmental Engineering\\
  Texas A\&M University\\
  College Station, TX 77843 \\
  \texttt{faxi.yuan@tamu.edu} \\
  \And
 Ali Mostafavi \\
  Urban Resilience.AI Lab\\
  Zachry Department of Civil and \\Environmental Engineering\\
  Texas A\&M University\\
  College Station, TX 77843 \\
  \texttt{amostafavi@civil.tamu.edu} \\
}

\begin{document}
\maketitle
\begin{abstract}
The objective of this study is to quantify community resilience based on fluctuations in the visits to various Point-of-Interest (POIs) locations during different stages of a crises. Visits to POIs is an important indicator of population activities and captures the combined effects of perturbations in people lifestyles, the built environment conditions, and the status of businesses. The study utilized digital trace data related to unique visits to POIs in the context of the 2017 Hurricane Harvey in Houston (Texas, USA) to examine the spatial patterns of impacts and total recovery efforts and utilized these measures in quantifying community resilience. The results showed that certain POI categories such as building materials and supplies dealers and grocery stores were the most resilient elements of the community compared to the other POI categories. On the other hand, categories such as medical facilities and entertainment were found to have lower resilience values. This result suggests that these categories were either not essential for community recovery or that the community was not able to access these services at normal levels immediately after the hurricane. In addition, the spatial analyses revealed that many areas in the community with lower levels of resilience (high systematic impact and long total recovery effort) experienced extensive flooding. However, some areas with low resilience were not flooded extensively, suggesting that spatial reach of the impacts goes beyond flooded areas. These results demonstrate the importance and value of the approach proposed in this study for quantifying and analysing community resilience patterns using digital trace/location intelligence data related to population activities. While this study focused on the Houston area and only analysed one natural hazard, the same approach can be applied to other communities and disaster contexts. By applying this approach, emergency managers and public officials can efficiently monitor the patterns of disaster impacts and recovery across different spatial areas and POI categories and also identify POI categories and areas of their community that need to be prioritized for resource allocation.
\end{abstract}

% keywords can be removed

\section{Introduction}
Communities affected by natural hazards are complex systems of interacting components (1,2,3,4). The complex system encompasses community residents with various personal lifestyles, the built environment providing various services to residents and businesses, and the hazards causing perturbations in the built environment, businesses, and daily lifestyles of residents (5). Quantifying community resilience is an integral part of effective response and resource allocation across different disaster phases (6,7). Existing studies have offered various approaches for quantifying community resilience. The disaster and community resilience literature has offered various approaches to assess the resilience of communities (8,9). These approaches, however, mainly focus on the resilience assessments related to individual components of the complex system (10). A growing body of studies (11,12,13) have focused on disaster impacts and recovery of the built environment and infrastructure systems. The resilience of the essential infrastructure sectors of communities such as hospitals (14,15), schools (16), businesses and supply chains (17,18), oil and gas (19) and grocery stores (20,21) have been examined in multiple prior studies. These studies have been helpful in understanding the vulnerability, reliability and recovery of the built environment during the disaster, which can inform the development of the repair and recovery plans. However, existing literature related to community resilience has paid relatively little attention to the interactions among populations, businesses, and the built environment (22,23). In fact, the built environment is a mean rather than an end in the resilience assessment (24). For example, while the absorptive capacity of hospitals is an important dimension in community resilience, people's access to healthcare services should also be considered during the disaster (25). This is particularly important for monitoring the effects and recovery of different segments of the community having varying levels of access to the essential services (26,27).

Another limitation is the dearth of empirical studies assessing community resilience based on data capturing the complex interactions in the nexus of populations, businesses, and the built environment. The majority of existing studies also fail to capture the effects of disaster-induced perturbations on population access to the critical services and population lifestyles due to disruptions in the built environment and businesses. Additionally, existing studies have mainly utilized two data types: surveys (28,22,29,30,31) and social media data (32,33,34,35,36,37). Surveys have two primary limitations for community resilience evaluations. First, surveys can only capture limited aspects of community resilience (such as household recovery) and their ability to capture the dynamic patterns of resilience and recovery is rather limited (38). Second, using surveys for collecting longitudinal and representative data from different locations across disaster-affected regions is challenging and often expensive. Hence, in addition to survey data, researchers have utilized social media platforms in community resilience assessment studies. Social media data has been applied to capturing societal disruptions (34), conducting rapid damage assessment (39,40,41), and sensing the dynamic situation of infrastructure services (42). Social media data, however, has limitations related to representativeness and can overlook certain demographic groups (43). In addition, only 1\% to 2\% of Twitter data contains location information (43). Considering these limitations in applying survey and social media data for community resilience studies, this study has elected for an entirely different and relatively new approach. This approach utilizes digital trace data related to visits to Point-of-Interest (POIs) in order to capture the dynamic interactions among people, disaster events and the built environment for community resilience assessment. Recent advancements in location intelligence technologies has led to the availability of the digital trace data related to POI visits. Digital trace data has been increasingly utilized in urban studies (e.g., mobility studies); however, the use of digital trace data in the context of disaster research has been rather limited (44,45). Moreover, digital trace data can be obtained and processed in a short time and help officials and disaster managers examine the dynamic patterns related to disaster impacts and recovery of the community during crises (46). 

In this study, we examine communities as complex systems and consider POI visits as indicators for capturing the dynamic state of communities. The state of a community as a complex system emerges based on the interactions among population lifestyles, state of businesses, and condition of the built environment (Figure 1). In a normal period without hazards, the system is in equilibrium. Perturbations in any component could alter the state of a community and such changes can be captured based on fluctuations in population activities (in particular, POI visits). Hence, POI visits capture the social, economic, and built environment state of communities. Accordingly, POI visits could be used as a holistic measure of performance to help quantify community resilience. This provides insight into the state of the complex system and enables spatial and temporal assessment of the disaster impact and recovery of the communities during and in the aftermath of disasters.  Thus, we utilized digital trace data related to POI visits in Houston, Texas in the context of 2017’s Hurricane Harvey. This enabled the quantification of community resilience and revealed the spatial patterns of community resilience. 

\begin{figure}[!ht]
\centering
\includegraphics[width=0.85\linewidth]{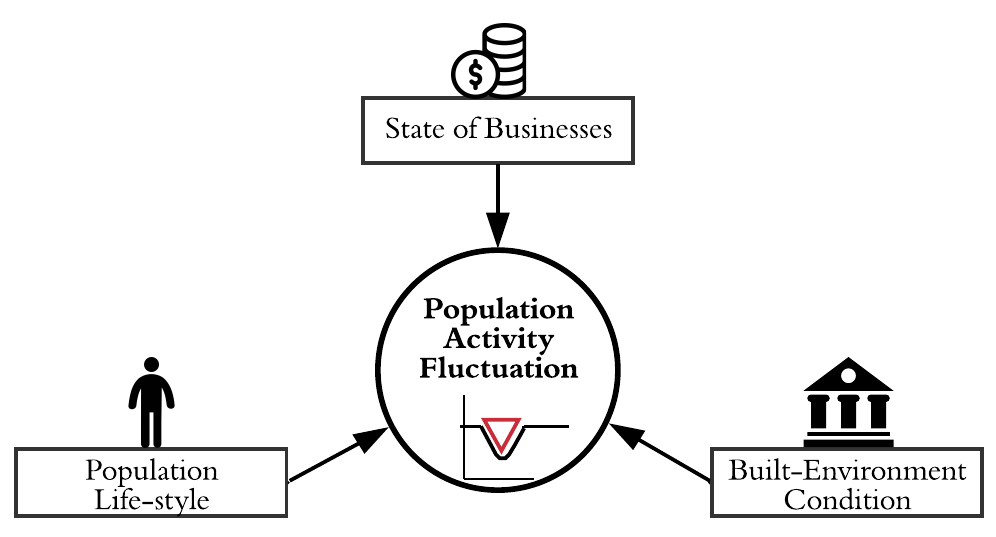}
\caption{Population activity fluctuations as an indicator of the state of community resilience}
\end{figure}

\section{Materials and Methods}

Aiming to quantify community resilience with fluctuations in POI visits, we established a methodological framework (Figure 2). Using the Hurricane Harvey’s impact on Houston as the case study, we implemented the framework by using the POI visit data from Aug. to Oct. 2017. We employed the resilience framework proposed in (47) by concentrating on systematic impact (i.e., the variances of the percent of POI visits before and after the disaster) and total recovery effort (i.e., the recovery time to the baseline of pre-disaster conditions). Community resilience was then quantified using a General Resilience (GR) metric for each POI category. 

\begin{figure}[ht]
\centering
\includegraphics[scale=0.25]{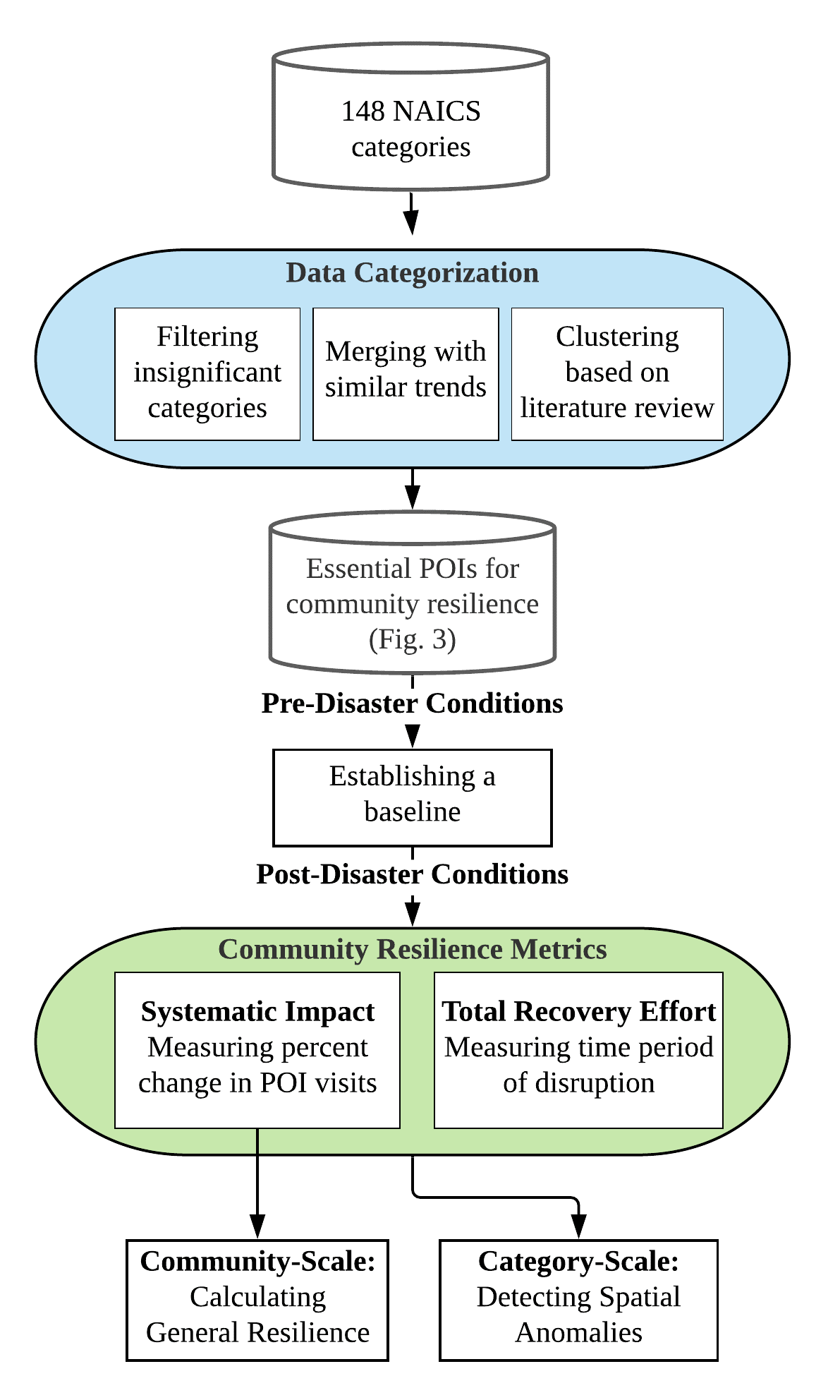}
\caption{Methodological framework for quantifying community resilience using POI visit data}
\end{figure}

\subsection{Data Description}
POI visit data was collected from the SafeGraph database. SafeGraph is a company that provides the "most accurate Point-of-Interest data and store location geofences for the U.S," and collects unique visit instances to physical locations in the community from anonymized cell-phone data. Hurricane Harvey made landfall on Aug. 25th as a category 4 hurricane. Houston experienced 27-54 inches of rain, shattering most known rainfall records (48).The data time frame contains the period before landfall (August 25), during the Hurricane (August 25 - August 31), and the weeks of immediate recovery period (September 1 – September 30). 

\subsection{Data Categorization}
Each month consisted of 55,537 distinct business entities (POIs) in Houston including daily visit count information and category type for each POI. Based on the North American Industry Classification System (NAICS), each POI was filtered from 148 pre-defined and fairly specific categories to 45 remaining categories. The literature review summarized in Table 1 discusses POIs essential to the main groups.
% \usepackage{multirow}
\begin{table}[]
\begin{tabular}{lll}
\hline
Essential POI Groups                        & References & Community   Resilience                                                   \\ \hline
\multirow{3}{*}{Emergency Preparedness}     & (49)       & Gasoline   station shortages                                             \\
                                            & (50)       & Increased   purchases of preparedness materials                          \\
                                            & (20,21)    & Greater attention   to grocery stores                                    \\ \hline
\multirow{2}{*}{Emergency Response}         & (14, 15)   & Resilience   frameworks for hospitals                                    \\
                                            & (51, 52)   & Organizational   structure of public order officials as first responders \\ \hline
\multirow{3}{*}{Recovery   Activity}        & (53)       & Increased   purchases of flood insurance for the recovery                \\
                                            & (54)       & Debris   clearance and immediate rebuilding                              \\
                                            & (30)       & Social   capital indices for resilience                                  \\ \hline
\multirow{2}{*}{Lifestyle   and Well-Being} & (55, 56)   & Business   and economic recovery of the community                        \\
                                            & (57)       & Addressing   well-being impacts from disaster impacts                    \\ \hline
\end{tabular}
\end{table}

Remaining POIs were merged based on their role in community resilience at different disaster stages (Figure 3). Next, we further classified these 16 categories into four main groups to explain their associations to the disaster impact: POIs essential for emergency preparedness, POIs essential for emergency response, POIs essential for lifestyle and well-being, and POIs essential for recovery activity.

\begin{figure}[ht]
\centering
\includegraphics[width=\linewidth]{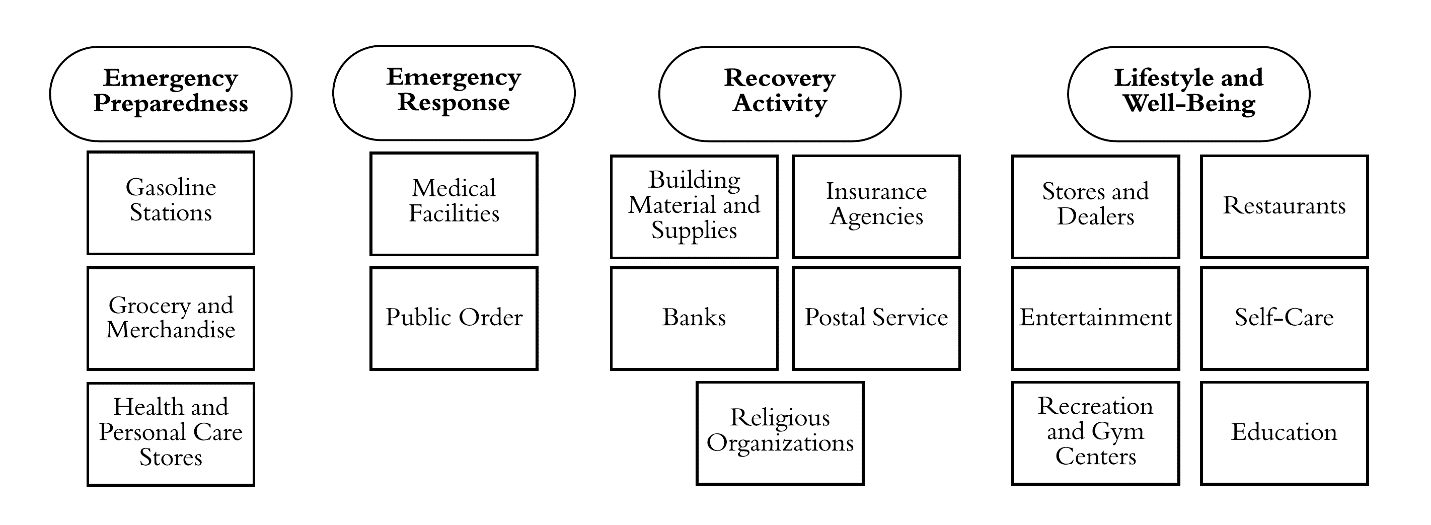}
\caption{POIs essential for community resilience during disasters}
\end{figure}

\subsection{Data Analysis}

\subsubsection{\textit{Establishing the Baseline for Pre-Disaster Conditions}}
To understand the pre-disaster trends in POI visits, we established a baseline for each of the 16 categories which served as a steady state of the community. POI visits in the first three weeks of Aug. were used to calculate the baseline except for the education category. Schools were just reopening in mid-to-late Aug. in the U.S., resulting in a steady increase in POI visits throughout this time period; and thus, the last three weeks of Oct. were instead utilized for the education baseline as this was when the trends in school visits were steadiest. To calculate the baseline, the number of visits on each weekday over the three selected weeks was averaged for every POI category individually. This process resulted in 16 category-specific baselines. Thereafter, we measured variances in POI visits compared with the baselines. 

\subsubsection{\textit{Calculating Percent Differences}}
The percent change indicates the increase or decrease from the baseline of daily visits for each of the 16 major POI categories. In Equation (1), the individual daily visit counts for each category were utilized as the “daily value”, and the aforementioned baseline points were used as the “baseline value”. When determining the actual percent change values throughout the time period, each weekday and weekend was compared to its corresponding baseline average. For instance, each Monday was compared to the Monday average, and each Sunday to the Sunday average. This was done to account for the differences in the daily POI visit patterns. Additionally, the 7-day rolling averages of these percent changes were calculated in order to alleviate daily anomalies and provide smoother trends. All analysis discussed in the following sections was performed on the 7-day rolling average values.
\begin{equation}
\label{eq:1}
Percent\ change=\frac{daily\ value-baseline\ value}{\left(daily\ value+baseline\ value\right)\ast\frac{1}{2}}
\end{equation}

\subsubsection{\textit{Elements of The Resilience Curve}}
Resilience curves (Figure 4) were determined for each POI Category and were confined by the transition point at disruption, TRNS(D), and the transition point signalling the new steady phase post disruption, TRNS (NS). For the purposes of this study, both values were taken at the baseline level of a 0 percent difference in POI visits. 

Measuring the Systematic Impact: After establishing the baseline and percent differences, each category’s systematic impact was quantified between Aug. 2017 and Sept. 2017 (Equation 1). Systematic impact refers to the lowest point reached in the resilience curve post disruption and is considered as the maximum level of disruption experienced by that POI category.
Measuring the Total Recovery Effort: We measured the total recovery effort, as the period of disruption for each POI category (i.e., the number of days taken for the daily percent change value to return to a zero after the initial point of disruption). This variable is the period from the TRNS(D) to TRNS(NS).
Calculating the General Resilience Values: We modified the general resilience (GR) metric developed by (47), and it is an integrated metric that considers each category’s systemic impact, slope ratio, recovery effort, and time averaged performance loss based on the resilience curve (Figure 4). We utilized the GR metric to quantify community resilience and further compared the resilience of various POIs in Houston in the context of Hurricane Harvey. The GR metric can be calculated for each POI category using Equation 2:

\begin{equation}
\label{eq:2}
GR=f(SI,SR,TAPL,\ RA)=\ SI\times SR\times{TAPL}^{-1}\times RA
\end{equation}

\begin{itemize}
\item Systematic Impact (SI): This value refers to the minimum point of percentage change of POI visits post disruption. It is the transition point signalling the shift from disruption to recovery and is the minimum point of percentage change of POI visits post system disruption.
\item Slope Ratio (SR): This value is taken as the slope of the recovery phase divided by the slope of the disruption phase. The slope of the disruption phase is the slope from the TRNS(D) value to the systematic impact. The slope of the recovery phase is the slope from the systematic impact to the TRNS(NS) value. These two phases are depicted in Figure 4. A higher slope ratio value indicates a quicker recovery compared to the level of disruption and its associated rapidity.
\item Time Average Performance Loss (TAPL): This value is the ratio of the area between the baseline level and the resilience curve and total recovery effort.
\item Recovery Ability (RA): This accounts for the extent to which the new steady state achieved post disruption compares to the steady state before the disruption. In this research, since the steady states were both taken as the 0\% difference baseline, this value is equal to 1 for all categories.
\end{itemize}

\begin{figure}[ht]
\centering
\includegraphics[scale=0.5]{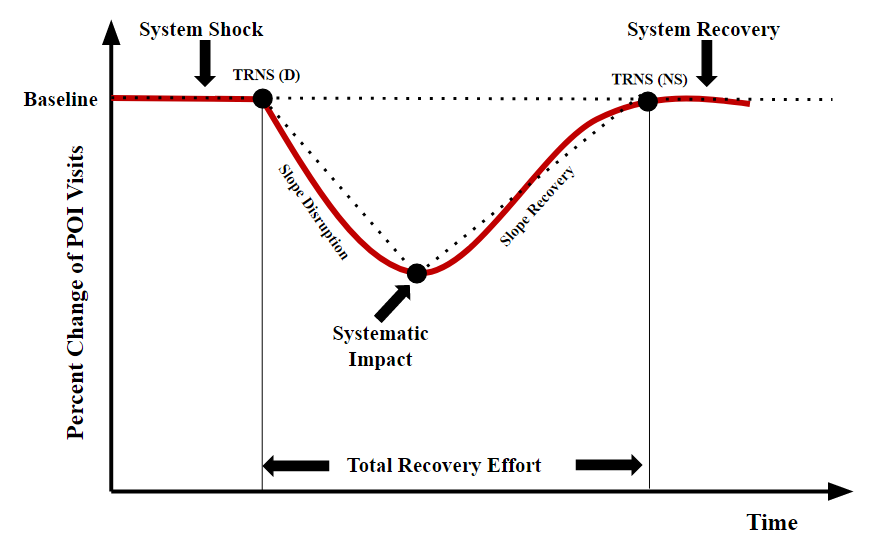}
\caption{Resilience curve of percentage change of POI visits due to system shock}
\end{figure}

\section{Percentage Change from the Baseline}

Raw daily visits were not used to directly calculate the systematic impact, slope ratio, and total recovery effort; instead, the 7-day rolling average was used to measure these community resilience aspects. As seen in the figures (e.g., gasoline stations in Figure 5), raw data values have a much greater negative percent change when compared to the 7-day rolling average. However, the 7-day rolling average minimizes the influence of random noise and fluctuations of the POI visits. Thus, the 7-day rolling average was used to calculate the systematic impact, total recovery effort, and ultimately the general resilience (GR) values.  

Figures 5-9 begin on Aug. 20th because the prior weeks were used to calculate each category’s percent baseline. As such, this region of each graph is relatively flat and near zero. This holds true for all categories except education where the last three weeks of Oct. were used for its baseline. Additionally, in multiple of the figures there is a notable one-day dip in the percent of raw daily visits on Sept. 4th, 2017 because of Labour Day. All non-essential government offices are closed on Labour Day, and many businesses are also closed or have modified hours. This directly explains the dramatic drops of POI visit data seen on Sept. 4th. Figures 5-9 illustrate the raw visit data in a cyan line; the associated 7-day rolling average in a dark blue line; the zero value in a black line; and the day of hurricane landfall (approximately on Aug. 25th) in a red line.

\subsection{POIs Essential for Emergency Preparedness} 
The POIs essential for emergency preparedness captures the various protective actions taken by residents to prepare for disasters. The POI categories include (1) gasoline stations, (2) grocery and merchandise, and (3) health and personal care stores. As shown in the percent change graphs, there were dramatic increases in the percent of raw daily visits right before Hurricane Harvey made landfall, and such an increase could be measurable shocks to the category. This implies that residents were refilling their cars and stockpiling groceries and personal products as preparedness measures. Although community residents are often advised to prepare for upcoming disasters, such sudden shocks may leave certain members of the community without proper resources. Compared to the other POI groups, this group generally experienced a low systematic impact, short total recovery effort, and GR values on the higher end of the spectrum.

\begin{figure}[ht]
\centering
\includegraphics[width=\linewidth]{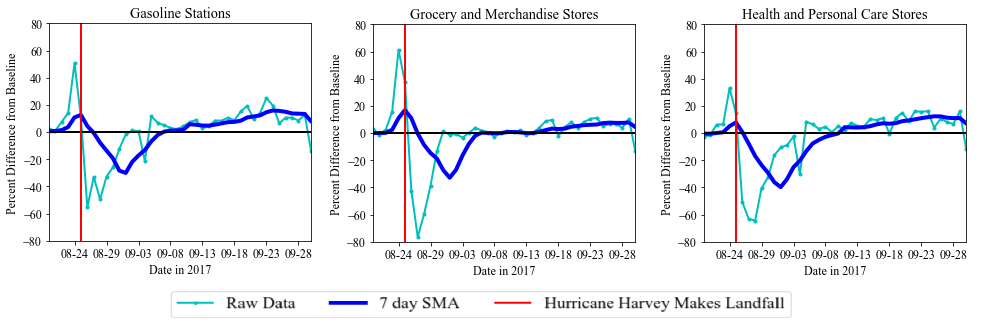}
\caption{Resilience curves for emergency preparedness POI categories}
\end{figure}

\subsubsection{\textit{Gasoline Stations}}
According to the raw daily visits, gasoline station visits experienced an increase of 50.82\% shortly before Hurricane Harvey made landfall. This result indicates that Houston residents could be filling up their tanks in preparation for the incoming hurricane. The 7-day rolling average had a systematic impact of a -30.06\%, a total recovery effort of 12 days, and a GR value of 1.87. 

\subsubsection{\textit{Grocery and Merchandise}}
Grocery stores include any independent food or meat markets as well as larger chain stores while merchandise stores include businesses such as Walmart. According to the raw daily visits, there was a sharp increase of 61.35\%. Similar to gas stations, this shows residents were taking preparedness actions; this time in the form of stockpiling food and supplies. The 7-day rolling average had a systematic impact of -32.87\%, a total recovery effort of 13 days, and a GR value of 2.32.

\subsubsection{\textit{Health and Personal Care Stores}}
Health and personal care stores include Walgreens and CVS pharmacies. According to the raw daily visits, the group experienced an increase of 33.06\%, although this was not as significant as the other two POIs essential for emergency preparedness. These facilities are not known to provide personal and prescribed medical remedies. The 7-day rolling average had a systematic impact of -39.77\%, a total recovery effort of 16 days, and a GR value of 1.62.

\subsection{POIs Essential for Emergency Response}
The POIs essential emergency response category provides medical and safety services to the community, which are most responsive during and in the immediate aftermath of the disaster (Figure 6). This category includes (1) medical facilities and (2) public order.

\begin{figure}[ht]
\centering
\includegraphics[scale=0.8]{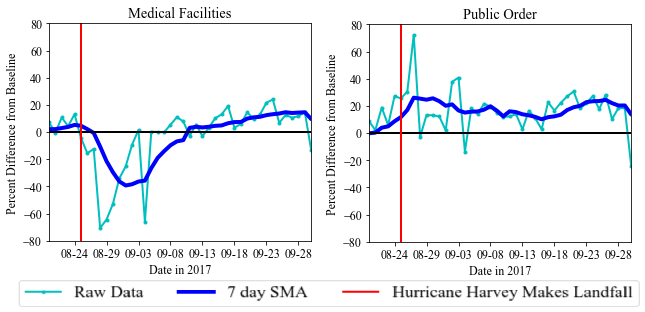}
\caption{Resilience curves for emergency response POI categories}
\end{figure}

\subsubsection{\textit{Medical Facilities}}
Medical facilities consist of offices of physicians and general medical and surgical hospitals. The 7-day rolling average had a systematic impact of -39.29\%, a total recovery effort of 16 days, and a GR value of 0.96. Though medical facilities had an average systematic impact and total recovery effort compared to other categories, it had the lowest GR value. This could indicate that people did not visit the physical locations of regular medical centres to seek medical support; however, they could have visited other temporarily available medical centres.

\subsubsection{\textit{Public Order}}
Public order consists of any justice, public order, and safety activity such as county precincts, fire departments, police departments. Unlike every other category, public order did not experience a negative percent change during Hurricane Harvey. Instead, the category had a staggering increase in raw daily visits, reaching a 72.27\% increase, and these above-baseline levels persisted until the beginning of Oct. The 7-day rolling average also saw a 26.05\% peak increase. This result could be due to the increased need and requests as more people would visit these locations looking for relief. Often, these facilities will also collaborate with and work alongside emergency responders to meet the immediate safety and medical needs of the community. Since there was no period of disruption, the systematic impact, total recovery effort and the GR value were not applicable to this POI category.

\subsection{POIs Essential for the Recovery Activity}
The POIs essential for the recovery activity category focuses on both the short-term and long-term recovery from the natural hazard such as rebuilding homes and managing finances (Figure 7). This category includes (1) insurance agencies, (2) building material and supplies, (3) banks, (4) postal service, and (5) religious services. An important finding was the gradual and sustained increase in raw daily visits during the recovery period in Sept.

\begin{figure}[ht]
\centering
\includegraphics[width=\linewidth]{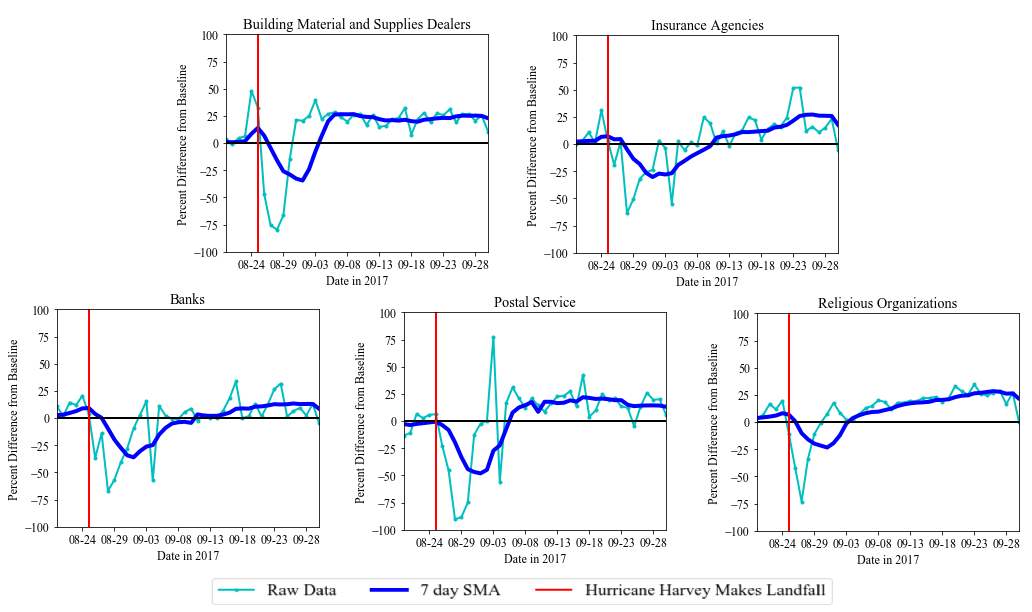}
\caption{Resilience curves for recovery activity POI categories}
\end{figure}

\subsubsection{\textit{Insurance Agencies}}
Insurance agencies refer to smaller, independent businesses as well as larger, corporate ones like Allstate Insurance and State Farm. The 7-day rolling average had a systematic impact of -30.34\%, a total recovery effort of 15 days, and a GR value of 1.08. Though the GR value is comparatively on the lower end of the scale, it does not account for the increased need displayed on the resilience curve. The gradual increase suggests that residents could be visiting locations to claim flood insurance for the disaster, which demonstrates a need to return to a state of normalcy. Residents could also be opening up new insurance accounts for future disasters.

\subsubsection{\textit{Building Material and Supplies}}
The building material and supplies include independent hardware and construction material stores as well as large chain stores such as Home Depot, Lowe’s, and Ace Hardware. The constant above-baseline percentages of the raw daily visits suggest that individual residents could still be rebuilding well after the event such as making structural repairs in the home. The 7-day rolling average had a systematic impact of -34.43\%, a total recovery effort of 11 days, and a GR value of 4.11. Building material and supplies had the highest GR value which suggests the immediate need to recover the home after a disaster.

\subsubsection{\textit{Banks}}
Banks include larger nationwide banks such as Bank of America and Chase and smaller, local banks. The 7-day rolling average had a systematic impact of -36.09\%, a total recovery effort of 16 days, and a GR value of 1.17. Banks had an average systematic impact and an average total recovery effort along with a low GR value. This could indicate that residents may not immediately need to attend a physical bank to access their finances and instead would rely on online services.

\subsubsection{\textit{Postal Service}}
This consists almost entirely of United States Postal Service facilities. According to the raw daily visits, postal service category experienced a 90.27\% decrease, the largest drop in raw daily visits. The 7-day rolling average had a systematic impact of -48.07\%, a total recovery effort of 13 days, and a GR value of 2.9, or the second highest GR value. The offset of the raw daily visits could indicate the high need for using the postal service.

\subsubsection{\textit{Religious Organizations}}
The 7-day rolling average had a systematic impact of -23.38\% and a total recovery effort of 9 days, which resulted in a GR value of 2.86. Although the intention of the visits cannot be confirmed, religious organizations are often used to house displaced residents and give out resources to those in need. This could explain the comparatively low systematic impact, low total recovery effort, and relatively high recovery effort.

\subsection{POIs Essential for Lifestyle and Well-being}
The POIs for lifestyle and well-being represent the economic and well-being needs of the community (Figure 8). The category includes: (1) stores and dealers, (2) restaurants, (3) entertainment, (4) self-care, and (5) recreation and gym centres, and (6) education. This category did not experience clear increases regarding raw daily visits staying below 18\%, which suggests that residents do not need and did not prioritize visiting these locations before the disaster. These categories often experienced the greatest systematic impact and longest total recovery effort compared to other POIs.

\begin{figure}[ht]
\centering
\includegraphics[width=\linewidth]{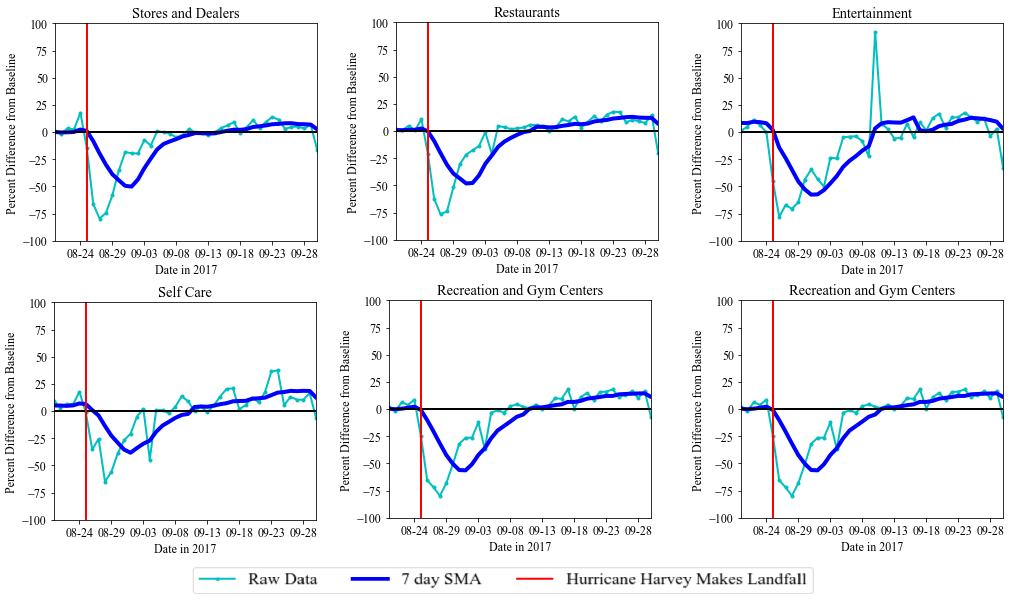}
\caption{Resilience curves for recovery activity POI categories}
\label{fig:congetsion}
\end{figure}

\subsubsection{\textit{Stores and Dealers}}
The stores and dealers include but are not limited to: clothing stores, sporting goods stores, book stores, electronic stores, and liquor stores, and also includes lessors of real estate such as storage facilities and shopping centres. The 7-day rolling average had a systematic impact of -50.14\%, a total recovery effort of 18 days, and a GR value of 1.62.

\subsubsection{\textit{Restaurants}}
This group refers to chain restaurants, bars, and beverage establishments. The 7-day rolling average had a systematic impact of -48.17\%, a total recovery effort of 17 days, and a GR value of 1.34.

\subsubsection{\textit{Entertainment}}
Entertainment includes locations such as museums, historical sites, and movie theatres, which are associated with places of leisure and social gatherings. During the recovery phase, one notable increase of raw daily visits was on Sunday, Sept. 10. This category includes NRG Stadium, the home of the Houston Texans football team, and upon inspection of the team’s 2017 schedule, it became clear that this spike coincides with a Texans’ home game. The 7-day rolling average had a systematic impact of -57.54\%, a total recovery effort of 17 days, and a GR value of 1.09.

\subsubsection{\textit{Self-Care}}
Self-care consists of beauty salons and studios, barbershops, and dentist offices, which address personal care not as immediate as medical services. The 7-day rolling average had a systematic impact of -38.29\%, a total recovery effort of 17 days, and a GR value of 1.47.

\subsubsection{\textit{Recreation and Gym Centres}}
This group refers to gyms, country clubs, golf courses, and other sports and fitness related centres. The 7-day rolling average had a systematic impact of -56.35\%, a total recovery effort of 18 days, and a GR value of 1.47.

\subsubsection{\textit{Education}}
Education consists of primary schools, secondary schools, and colleges. The 7-day rolling average had a systematic impact of -75.94\%, a total recovery effort of 21 days, and a GR value of 1.25. Education had the greatest systematic impact and longest total recovery effort with a very low GR value. This makes sense as community residents are more concerned with meeting their immediate needs such as water, food, and shelter rather than education. Education facilities are also very large and require lots of care to restore. 

\section{Community- Scale Analysis: Resilience Metrics of POI Categories}

The dynamic interactions between the built-environment, individual households, and the disaster event are represented by POI visits due to Hurricane Harvey. The range of systematic impact, total recovery effort, and GR values demonstrate how certain POI categories could be considered more resilient comparatively within a community. For example, religious organizations experienced the least systematic impact at -30.06\% and the shortest total recovery effort at 9 days whereas education experienced the greatest systematic impact at -75.94\% and the longest total recovery effort at 21 days (Figure 9 and Figure 10). 

\begin{figure}[ht]
\centering
\includegraphics[scale=0.65]{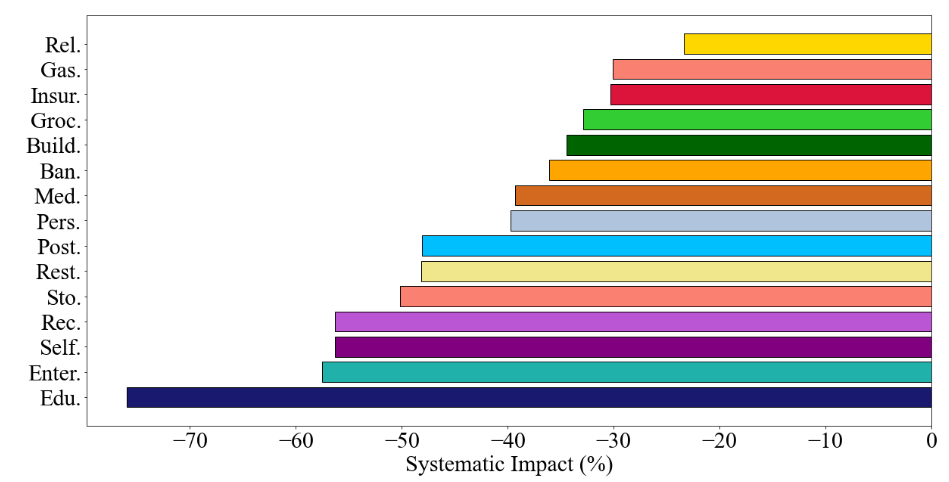}
\caption{Systematic impact value for every POI category, ranked from smallest to largest}
\end{figure}

\begin{figure}[ht]
\centering
\includegraphics[scale=0.65]{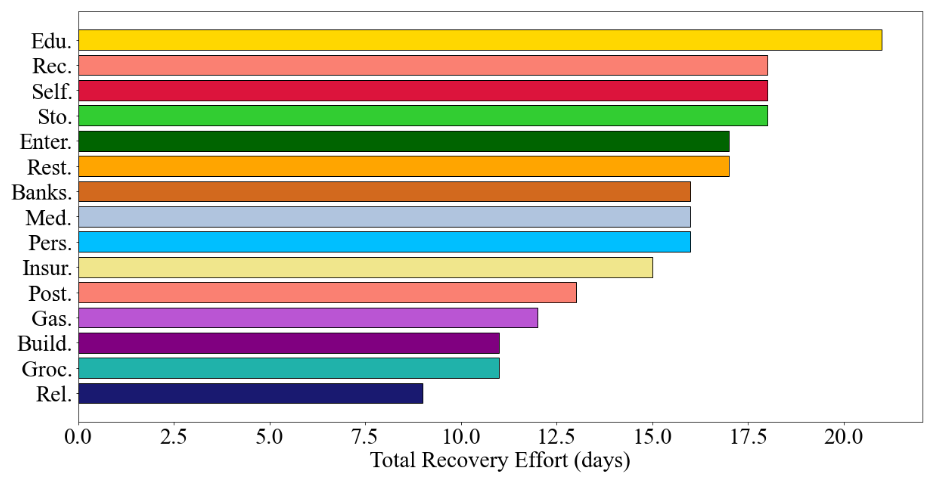}
\caption{Total recovery effort value for every POI category, ranked from largest to smallest}
\end{figure}

The general trend in Figure 11 depicts an overall direct relationship between higher GR values and higher slope ratios. A higher slope ratio indicates a higher recovery slope than disruption slope, which is a component necessary for a higher GR value. Outliers of this trend have these differences attributed to the time averaged performance loss (TAPL) discussed previously. The higher the GR value, the more resilient that POI category can be considered. Thus, building materials and supplies and grocery and merchandise are considered to be more resilient categories while entertainment and medical facilities are considered to be less resilient categories. 

\begin{figure}[ht]
\centering
\includegraphics[scale=0.65]{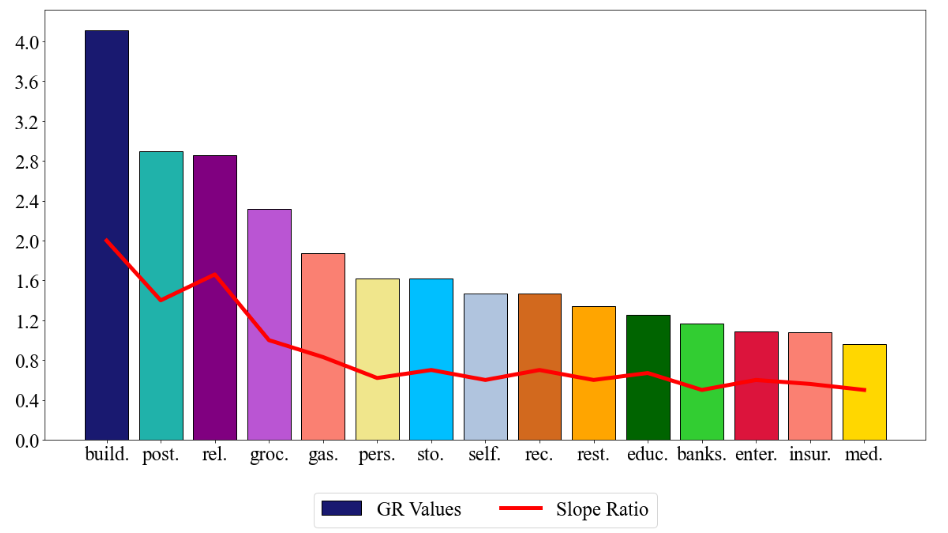}
\caption{General resilience (GR) metric values with slope ratios}
\end{figure}

One category has been identified per group to provide a more in-depth discussion regarding how the resilience metrics capture a different area of community resilience. Certain categories were considered to be comparatively more resilient. For example, the grocery and merchandise category in the emergency preparedness group captures the preparedness actions residents took before Hurricane Harvey, as explained by this category’s high peak in visitations just before landfall. This category also has one of the highest GR values which could be attributed to it having the 2nd shortest total recovery effort. This result suggests that the grocery and merchandise category as a whole was only negatively impacted for a relatively short period before bouncing back to normal visitation levels. The building material and supplies category in the recovery activity group captures the rebuilding effort in after Hurricane Harvey. It has the highest GR value, mainly because the slope of its recovery phase was twice that of its disruption phase. This result suggests that residents needed this service as quickly as possible and at higher rates after Hurricane Harvey. Thus, the community had a great dependence on the grocery and merchandise category and the building material and supplies category which could have contributed to their high resilience. Moreover, community leaders must also take great care in ensuring and maintaining the functionality of these POI categories.

On the other hand, certain categories could be considered to be less resilient due to a lack of accessibility to the category or less dependence on the category. The medical facilities in the emergency response group partially captures the emergency support utilized after the hurricane occurred. Notably, this category had the lowest GR value of all as the category’s disruption phase slope was twice that of its recovery phase. In addition to the sharp reduction of visitations to medical facilities during the hurricane, it also took much longer to recover to baseline levels. Though residents could have been receiving primary medical care from other services such as emergency responders, it is important to note the lack of visitation to this critical category due to road inundations and disrupted access. Lastly, entertainment category in the lifestyle and well-being can reveal when the community returned back to a state of normalcy after Hurricane Harvey. This category had one of the lowest GR values, suggesting that it is less resilient than most other categories. It also had one of the longest total recovery efforts, the second highest systematic impact value, and a low slope ratio of 0.6, all of which further resulted in the low GR value. This suggest that the community only began to return to entertainment facilities once their other basic needs such as food, supplies, and shelter were met.

\section{Category-Scale Analysis: Spatial Disparities of Community Resilience}
This section introduces the examination of the spatial disparities of community resilience. Taking the POI category of building material and supplies as an example, we illustrate the spatial patterns of systematic impact and total recovery effort for different census tract areas in Figures 12 and 13. The maps divide the values into five 20-percentile division from greatest to least disaster impact (ranging from dark red to dark blue colours in Figures 12 and 13). The systematic impact was capped at a maximum value of zero while the total recovery effort was set at a minimum of zero. Higher negative systematic impact values represent greater disaster impact while longer positive total recovery effort values may also indicate greater disaster impact. Blank areas represent census tracts with no POIs from the building material and supplies. Based on observation of the maps (indicated by the dotted lines), the systematic impact and total recovery effort values generally overlap. In other words, greater systematic impact appears simultaneously with longer recovery effort, which has been confirmed with the Pearson Correlation value of -0.77. 

When comparing Figures 12 and 13 to the Hurricane Harvey flood scenario in Houston, the results reveal the relationship between the POI visits and flooding regions. For example, the mid-western area, overlaps with an area known for frequent flooding. Thus, the higher systematic impact and longer total recovery effort can be attributed to the increased flooding. In contrast, the northern red area did not overlap with a high flooding area. This result reveal that the spatial reach of impacts extending beyond flooded areas. In addition, the spatial distribution of POI categories can show areas of lower resilience which cannot be detected solely from flooding. Hence, the systematic impact and total recovery effort can be used to further understand the expected intensity and period of disruption caused by hazard events.

\begin{figure}[ht]
\centering
\includegraphics[width=\linewidth]{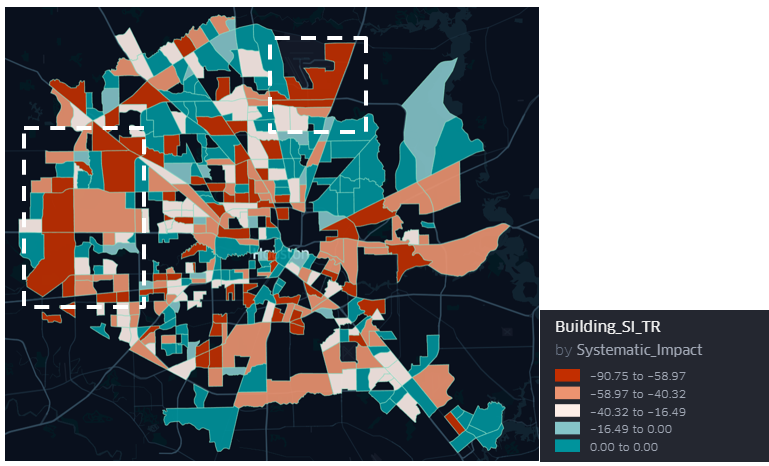}
\caption{Spatial distribution for systematic impact of building material and supplies}
\end{figure}

\begin{figure}[ht]
\centering
\includegraphics[width=\linewidth]{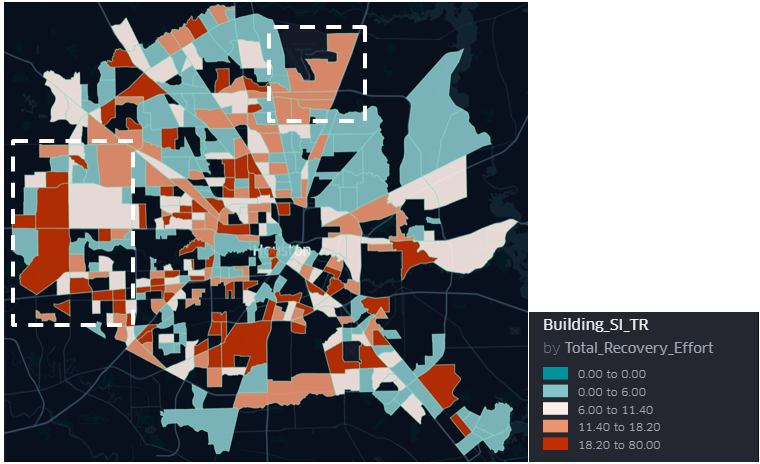}
\caption{Spatial distribution for total recovery effort of building material and supplies}
\end{figure}

\section{Limitations}
Given the novelty of this type of location-intelligence in academic research, it is important to discuss the possible limitations of the dataset along with guidelines for future use of the data. The first limitation is the demographics associated with mobility ownership and ultimately the collection of the location-intelligence data (59). The second limitation was the selection of the baseline. The research established a return to normalcy as reaching “pre-disaster” conditions of POI visits (60,61). It is important to consider the dynamic mobility patterns of a system to establish an accurate baseline, which is used for the comparison with the state after the shock strikes the system. The third limitation is the external factors influencing the data, such as major community events or celebrations which would influence the mobility of the system. An example of this was Labour Day, which coincided with drops in the majority of the categories. To mitigate this, the 7-day rolling average was used to calculate the systematic impact, slope ratio, and total recovery effort, and ultimately the final GR calculations.  

\section{Concluding Remarks}
This paper uses location intelligence data related to visits to Point-of-Interest (POIs) as an indicator for assessing the state of complex sociotechnical system of communities for resilience assessment. Natural hazards cause perturbation to the complex system of communities including the residents, built environment and state of the businesses. The aggregated measure of visits to POIs enables the understanding in interaction among these different components of communities as complex systems. The adopted approach in this study shows how researchers and practitioners are now able to devise inclusive methods for examining community resilience thanks to the growing availability of digital trace data. Implementing digital trace data, unlike survey data, is rather quick and could cover large areas for assessing the resilience of communities. In addition, the data are directly related to how residents interact with the built environment and business services which makes this approach a better choice compared to that by social media data. Hence, this study illustrates that using digit trace data (i.e., POI visit) complements the current resilience assessment approaches and helps provide a broader view of the interacting complex sociotechnical systems of the communities.

In addition, this study provides a novel quantitative approach to community resilience assessment. We quantify community resilience by examining unexpected shocks to essential facilities within the community. In particular, we focus on the variations of POI visits before and after the landfall of Hurricane Harvey in Houston. This study measures the systemic impact, recovery effort, and general resilience score of community resilience with the fluctuations of POI visits to four POI groups, including emergency preparedness, emergency response, recovery activity, and lifestyle and well-being.  On a community-wide scale, certain categories are more or less resilient compared to others. For instance, grocery and merchandise had a comparatively high GR value which could be due to the great dependence of community residents for securing food and supplies after the disaster. Interestingly, building material and supplies had the highest GR value which could signal the essential needs of rebuilding efforts after Hurricane Harvey. On the other side, medical facilities had the lowest GR values possibly due to lack of accessibility to these categories while the entertainment category had a low GR value possibly due to lack of need for the service. 

Assessing the resilience of the four essential aspects of community resilience brings a better empirical understanding in the way disaster researchers view unexpected shocks to systems. This approach provides the necessary information to bridging the gap between the physical and social impacts of the disaster. The findings also benefit disaster practitioners and emergency responders to prioritize the restoration and investment of different POI categories as well as visualize the intensity and period of disruption. Community leaders and emergency planners should take notice of POI categories with extremely high and extremely low resilience metric values. For example, services catered to basic needs such as food, supplies, and shelter generally had higher GR values; thus, community leaders and emergency planners should ensure that these related POI categories are functionally operational throughout the disaster. However, services catered to entertainment and shopping had lower GR values; thus, greater care must be given to the types of categories to ensure the long-term economic and well-begin stability of the community.

By comparing the systematic impact, total recovery effort, and general resilience (GR) of the POI categories, this research is able to provide resilience metrics for future application. On a category level, the spatial visualizations show census tracts of high and low disaster impact which could not be captured solely from flooding maps. For example, some areas of high disaster impact would not overlap with pre-established areas of frequent flooding. The combination of the community-wide scale and category-scale analysis provides a thorough understanding of the interactions between community residents and the built-environment.

%\bibliographystyle{unsrt}  

\section*{Acknowledgements}
The authors would like to acknowledge the funding support from the National Science Foundation (CAREER) under grant number 1846069. The authors would also like to acknowledge SafeGraph for providing the Point-of-Interest (POI) data. Any opinions, findings, conclusions, or recommendations expressed in this research are those of the authors and do not necessarily reflect the view of the funding agencies.

\section*{References}
1.  Li, Q., Dong, S., Mostafavi, A. (2019). “Modeling of inter-organizational coordination dynamics in resilience planning of infrastructure systems: A multilayer network simulation framework.” PLOS ONE, 14(11): e0224522.

2.  de Almeida, B. A., and Mostafavi, A. (2016). “Resilience of infrastructure systems to sea-level rise in coastal areas: Impacts, adaptation measures, and implementation challenges.” Sustainability (Switzerland).

3.  Mostafavi, A., Ganapati, N. E., Nazarnia, H., Pradhananga, N., and Khanal, R. (2018). Adaptive capacity under chronic stressors: Assessment of water infrastructure resilience in 2015 Nepalese earthquake using a system approach. Natural Hazards Review, 19(1), 05017006.

4.  Peacock, W. G., and Ragsdale, A. K. (1997). “Social systems, ecological networks and disasters: Toward a socio-political ecology of disasters.” Hurricane Andrew: Ethnicity, gender, and the sociology of disasters, Routledge, New York, 20–35.

5.  Rasoulkhani, K., Mostafavi, A., Presa, M., and Batouli, M. (2020). “Resilience planning in hazards-humans-infrastructure nexus:A multi-agent simulation for exploratory assessment of coastal water supply infrastructure adaptation to sea-level rise.” Environmental Modelling and Software, Elsevier Ltd, 125(January), 104636.

6.  Ramirez-marquez, J. E., Rocco, C. M., Barker, K., and Moronta, J. (2018). “Quantifying the resilience of community structures in networks.” Elsevier Ltd, 169(August 2017), 466–474.

7.  Yu, J. (2019). “Quantifying Community Resilience Using Hierarchical Bayesian Kernel Methods: A Case Study on Recovery from Power Outages.” 39(9), 1930–1948.

8.  Dong, S., Wang, H., Mostafavi, A., and Gao, J. (2019). “Robust component: A robustness measure that incorporates access to critical facilities under disruptions.” Journal of the Royal Society Interface, 16(157).

9.  Davis, C. A., Mostafavi, A., and Wang, H. (2018). “Establishing Characteristics to Operationalize Resilience for Lifeline Systems.” Natural Hazards Review, 19(4).

10. Applied Technology Council. (2016). Critical Assessment of Lifeline System Performance: Understanding Societal Needs in Disaster Recovery.

11. Batouli, M., and Mostafavi, A. (2018). “Multiagent Simulation for Complex Adaptive Modeling of Road Infrastructure Resilience to Sea-Level Rise.” Computer-Aided Civil and Infrastructure Engineering.

12. Guikema, S., and Nateghi, R. (2018). “Modeling Power Outage Risk From Natural Hazards.” 1(October), 1–28.

13. O’Rourke, T. (2007). “Critical infrastructure, interdependencies, and resilience.” BRIDGE-WASHINGTON-NATIONAL ACADEMY OF ENGINEERING.

14. Cimellaro, G. P., Reinhorn, A. M., Bruneau, M., Paolo, G., Reinhorn, A. M., and Seismic, M. B. (2010). “Seismic resilience of a hospital system.” 2479.

15. Zhong, S., Clark, M., Hou, X., Zang, Y., and Fitzgerald, G. (2014). “Development of hospital disaster resilience: conceptual framework and potential measurement.” (1), 930–938.

16. Baum, N. L., Rotter, B., Reidler, E., and Brom, D. (2009). “Building Resilience in Schools in the Wake of Hurricane Katrina.” Journal of Child \& Adolescent Trauma, 2(1), 62–70.

17. Aghababaei, M., Koliou, M., Watson, M., and Xiao, Y. (2020). “Quantifying post-disaster business recovery through Bayesian methods.” Structure and Infrastructure Engineering, Taylor \& Francis, 0(0), 1–19.

18. Middleton, E. J. T., and Latty, T. (2016). “Resilience in social insect infrastructure systems.” Journal of the Royal Society Interface.

19. Comes, T., and Van De Walle, B. (2014). “Measuring disaster resilience: The impact of hurricane sandy on critical infrastructure systems.” ISCRAM 2014 Conference Proceedings - 11th International Conference on Information Systems for Crisis Response and Management.

20. Singh-peterson, L., and Lawrence, G. (2015). “Insights into community vulnerability and resilience following natural disasters: perspectives with food retailers in Northern NSW, Australia.” Local Environment, Taylor \& Francis, 0(0), 1–14.

21. Spiegler, V. L. M., Potter, A. T., Naim, M. M., Towill, D. R., Potter, A. T., Naim, M. M., and The, D. R. T. (2016). “The value of nonlinear control theory in investigating the underlying dynamics and resilience of a grocery supply chain.” International Journal of Production Research, Taylor \& Francis, 7543, 1–22.

22. Esmalian, A., Dong, S., Coleman, N., and Mostafavi, A. (2019). “Determinants of risk disparity due to infrastructure service losses in disasters: a household service gap model.” Risk Analysis.

23. Mostafavi, A., and Ganapati, N. E. (2019). “Toward Convergence Disaster Research: Building Integrative Theories Using Simulation.” Risk Analysis.

24. Weilant, S., Strong, A., and Miller, B. (2019). “Incorporating Resilience into Transportation Planning and Assessment.” Incorporating Resilience into Transportation Planning and Assessment.

25. Dong, S., Esmalian, A., Farahmand, H., and Mostafavi, A. (2020). “An integrated physical-social analysis of disrupted access to critical facilities and community service-loss tolerance in urban flooding.” Computers, Environment and Urban Systems, 80.

26. Coleman, N., Esmalian, A., and Mostafavi, A. (2020). “Anatomy of Susceptibility for Shelter-in-Place Households Facing Infrastructure Service Disruptions Caused by Natural Hazards.” International Journal of Disaster Risk Reduction, Elsevier BV, 101875.

27. Coleman, N., Esmalian, A., and Mostafavi, A. (2020). “Anatomy of Susceptibility for Shelter-in-Place Households Facing Infrastructure Service Disruptions Caused by Natural Hazards.” International Journal of Disaster Risk Reduction.

28. Berlemann, M. (2016). “Does hurricane risk affect individual well-being? Empirical evidence on the indirect effects of natural disasters.” Ecological Economics, 124.

29. Morss, R. E., Mulder, K. J., Lazo, J. K., and Demuth, J. L. (2016). “How do people perceive, understand, and anticipate responding to flash flood risks and warnings? Results from a public survey in Boulder, Colorado, USA.” Journal of Hydrology, 541.

30. Sherrieb, K., Norris, F. H., and Galea, S. (2010). “Measuring Capacities for Community Resilience.” Social Indicators Research, 99(2).

31. Webb, G. R., Tierney, K. J., and Dahlhamer, J. M. (2002). “Predicting long-term business recovery from disaster: A comparison of the Loma Prieta earthquake and Hurricane Andrew.” Environmental Hazards.

32. Wang, Q., and Taylor, J. E. (2014). “Quantifying human mobility perturbation and resilience in hurricane sandy.” PLoS ONE, 9(11).

33. Zhai, W., Peng, Z. R., \& Yuan, F. (2020). “Examine the effects of neighborhood equity on disaster situational awareness: Harness machine learning and geotagged Twitter data.” International Journal of Disaster Risk Reduction, 101611.

34. Zhang, C., Fan, C., Yao, W., Hu, X., and Mostafavi, A. (2019). “Social media for intelligent public information and warning in disasters: An interdisciplinary review.” International Journal of Information Management, Elsevier, 49(April), 190–207.

35. Zhang, C., Yao, W., Yang, Y., Huang, R., and Mostafavi, A. (2020). “Semiautomated social media analytics for sensing societal impacts due to community disruptions during disasters.” 1, 1–18.

36. Yuan, F., Liu, R., Mao, L., \& Li, M. (2021). “Internet of people enabled framework for evaluating performance loss and resilience of urban critical infrastructures.” Safety Science, 134, 105079. 

37. Yuan, F., Li, M., \& Liu, R. (2020). “Understanding the evolutions of public responses using social media: Hurricane Matthew case study.” International Journal of Disaster Risk Reduction, 51, 101798.

38. Rasoulkhani, K., and Mostafavi, A. (2018). “Resilience as an emergent property of human-infrastructure dynamics: A multi-agent simulation model for characterizing regime shifts and tipping point behaviors in infrastructure systems.” PLoS ONE.

39. Yuan, F., and Liu, R. (2020). “Mining Social Media Data for Rapid Damage Assessment during Hurricane Matthew: Feasibility Study.” Journal of Computing in Civil Engineering, 34(3), 1–14.

40. Yuan, F., \& Liu, R. (2019). “Identifying damage-related social media data during Hurricane Matthew: A machine learning approach.” In Computing in Civil Engineering 2019: Visualization, Information Modeling, and Simulation (pp. 207-214). Reston, VA: American Society of Civil Engineers.

41. Yuan, F., \& Liu, R. (2018). “Feasibility study of using crowdsourcing to identify critical affected areas for rapid damage assessment: Hurricane Matthew case study.” International journal of disaster risk reduction, 28, 758-767.

42. Fan, C. (2019). “A graph-based method for social sensing of infrastructure disruptions in disasters.” (May), 1055–1070.

43. Fan, C., Esparza, M., Dargin, J., Wu, F., Oztekin, B., and Mostafavi, A. (2020). “Spatial biases in crowdsourced data: Social media content attention concentrates on populous areas in disasters.” Computers, Environment and Urban Systems.

44. Fan, C., Zhang, C., Yahja, A., \& Mostafavi, A. (2019). Disaster City Digital Twin: A vision for integrating artificial and human intelligence for disaster management. International Journal of Information Management, 102049.

45. Yabe, T., Tsubouchi, K., Fujiwara, N., Sekimoto, Y., and Ukkusuri, S. V. (2020). “Understanding post-disaster population recovery patterns.” Journal of the Royal Society Interface, 17(163).

46. Yabe, T., Zhang, Y., and Ukkusuri, S. (2020). “Quantifying the Economic Impact of Extreme Shocks on Businesses using Human Mobility Data: a Bayesian Causal Inference Approach.” arXiv:2004.11121.

47. Vugrin, E. D., Warren, D. E., Ehlen, M. A., \& Camphouse, R. C. (2010). A framework for assessing the resilience of infrastructure and economic systems. In Sustainable and resilient critical infrastructure systems (pp. 77-116). Springer, Berlin, Heidelberg.

48. US Department of Commerce NOAA. Major Hurricane Harvey - August 25-29, 2017 [Internet]. National Weather Service. NOAA's National Weather Service; 2019. Available from: https://www.weather.gov/crp/hurricane\_harvey

49. Khare, A., He, Q., and Batta, R. (2020). “Predicting gasoline shortage during disasters using social media.” OR Spectrum, 42(3).

50. Beatty, T. K. M., Shimshack, J. P., and Volpe, R. J. (2019). “Disaster preparedness and disaster response: Evidence from sales of emergency supplies before and after hurricanes.” Journal of the Association of Environmental and Resource Economists, 6(4).

51. Adams, T. M., and Stewart, L. D. (2015). “Chaos Theory and Organizational Crisis: A Theoretical Analysis of the Challenges Faced by the New Orleans Police Department During Hurricane Katrina.” Public Organization Review, 15(3).

52. Murphy, H. (2019). “Journal of Public Administration Research and Theory Police as Emergency Responders: Organizational or Personal Resilience?” Journal of Public Administration Research and Theory, 29(4).

53. Kousky, C. (2017). “Disasters as Learning Experiences or Disasters as Policy Opportunities? Examining Flood Insurance Purchases after Hurricanes.” Risk Analysis, 37(3).

54. Schwab, J. C. (2010). “Hazard Mitigation: Integrating Best Practices into Planning.” (560), 156.

55. Corey, C. M., and Deitch, E. A. (2011). “Factors affecting business recovery immediately after Hurricane Katrina.” Journal of Contingencies and Crisis Management, 19(3).

56. Marshall, M. I., and Schrank, H. L. (2014). “Small business disaster recovery: A research framework.” Natural Hazards.

57. Dargin, J. S., and Mostafavi, A. (2020). “Human-centric infrastructure resilience: Uncovering well-being risk disparity due to infrastructure disruptions in disasters.” PLoS ONE, 15(6).

58. Nan, C., and Sansavini, G. (2017). “A quantitative method for assessing resilience of interdependent infrastructures.” Reliability Engineering and System Safety, 157.

59. Wesolowski, A., Eagle, N., Noor, A. M., Snow, R. W., and Buckee, C. O. (2013). “The impact of biases in mobile phone ownership on estimates of human mobility.” Journal of the Royal Society Interface, 10(81).

60. Chang, S. E., and Chang, S. E. (2016). “Socioeconomic Impacts of Infrastructure Disruptions.” Oxford Research Encyclopedia of Natural Hazard Science.

61. McDaniels, T. L., Chang, S. E., Hawkins, D., Chew, G., and Longstaff, H. (2015). “Towards disaster-resilient cities: an approach for setting priorities in infrastructure mitigation efforts.” Environment Systems and Decisions, 35(2).

\section*{Supplemental Information}
\appendix
\counterwithin{figure}{section}

\section{Categorization of POIs}
The following section lists the U.S. Department Codes (NAICS) of the filtered data analysis. This includes 45 NAICS into 16 categories and 4 main groups.
\begin{itemize}
   \item I.    Emergency Preparedness Sector
   \begin{itemize}
     \item 1. Gasoline Stations ()
        \begin{itemize}
            \item a.	Gasoline Stations
        \end{itemize}
     \item 2.	Grocery and Merchandise ()
        \begin{itemize}
            \item a.	Grocery Stores
            \item b.	General Merchandise Stores, including Warehouse Clubs and Supercenters
        \end{itemize}
     \item 3.	Health and Personal Care Stores ()
        \begin{itemize}
            \item a.	Health and Personal Care Stores
        \end{itemize}
   \end{itemize}
   \item II.	Emergency Response Sector
   \begin{itemize}
     \item 4.	Medical Facilities ()
        \begin{itemize}
            \item a.	Offices of Physicians
            \item b.	General Medical and Surgical Hospitals
        \end{itemize}
     \item 5.	Public Order ()
        \begin{itemize}
            \item a.	Justice, Public Order, and Safety Activities
        \end{itemize}
   \end{itemize}
   \item III.  Recovery Activity Sector
   \begin{itemize}
     \item 6.	Building Material and Supplies ()
        \begin{itemize}
            \item a.	Building Material and Supplies Dealers
        \end{itemize}
     \item 7.	Insurance Agencies ()
        \begin{itemize}
            \item a.	Agencies, Brokerages, and Other Insurance Related Activities
        \end{itemize}
    \item 8.	Banks ()
        \begin{itemize}
            \item a.	Depository Credit Intermediation
        \end{itemize}
    \item 9.	Postal Service ()
        \begin{itemize}
            \item a.	Postal Service
        \end{itemize}
    \item 10.	Religious Organizations ()
        \begin{itemize}
            \item a.	Religious Organizations
        \end{itemize}
   \end{itemize}
    \item IV.	Lifestyle and Well-Being Sector
   \begin{itemize}
     \item 11.	Stores and Dealers ()
        \begin{itemize}
            \item a.	Clothing Stores
            \item b.	Used Merchandise Stores
            \item c.	Sporting Goods, Hobby, and Musical Instrument Stores
            \item d.	Specialty Food Store
            \item e.	Spectator Sports
            \item f.	Shoe Stores
            \item g.	Department Stores
            \item h.	Office Supplies, Stationery, and Gift Stores
            \item i.	Other Miscellaneous Store Retailers
            \item j.	Beer, Wine, and Liquor Stores
            \item k.	Direct Selling Establishments
            \item l.	Drinking Places (Alcoholic Beverages)
            \item m.	Electronics and Appliance Stores
            \item n.	Florists
            \item o.	Home Furnishings Stores
            \item p.	Furniture Stores
            \item q.	Jewelry, Luggage, and Leather Goods Stores
            \item r.	Automotive Parts, Accessories, and Tire Stores
            \item s.	Book Stores and News Dealers
            \item t.	Other Motor Vehicle Dealers
            \item u.	Automobile Dealers
            \item v.	Lessors of Real Estate
        \end{itemize}
     \item 12.	Restaurants ()
        \begin{itemize}
            \item a.	Restaurants and Other Eating Places
        \end{itemize}
    \item 13.	Entertainment ()
        \begin{itemize}
            \item a.	Museums, Historical Sites, and Similar Institutions
            \item b.	Motion Picture and Video Industries
            \item c.	Promoters of Performing Arts, Sports, and Similar Events
            \item d.	Amusement Parks and Arcades
        \end{itemize}
    \item 14.	Self-Care ()
        \begin{itemize}
            \item a.	Offices of Dentists
            \item b.	Personal Care Services
        \end{itemize}
    \item 15.	Recreation and Gym Centers	()
        \begin{itemize}
            \item a.	Other Amusement and Recreation Industries
        \end{itemize}
    \item 16.	Education ()
        \begin{itemize}
            \item a.	Elementary and Secondary Schools
            \item b.	Colleges, Universities, and Professional Schools
            \item c.	Junior Colleges
        \end{itemize}
   \end{itemize}
\end{itemize}

\section{Relevant Tabular Information about POI Categories}
We examined the raw daily visits through (1) hikes before the hurricane landfall, (2) maximum drops after the hurricane landfall, (3) and peaks during the system recovery. The exact values are summarized in the Supplemental Information. Throughout these results, hike refers to a sudden increase in raw daily visits just before Harvey made landfall, maximum drop refers to the highest negative percent change value of raw daily visits; and peak refers to the highest positive percent change after the raw daily visits returned/ passed the baseline levels during the system recovery. Table B1 presents the data information of the raw daily visits of the percent change graphs for the categories. Table B2 presents the average, median, and standard deviation of the systematic impact and total recovery effort for the POI categories. 
% \usepackage{multirow}
\begin{table}[]
\renewcommand\thetable{B1} 
   \centering
   \caption{Raw Daily Visits of POI Categories from the Percent Change Graphs}
\begin{tabular}{|l|l|l|l|l|l|l|}
\hline
\multirow{2}{*}{POI categories}                                            & \multicolumn{2}{l|}{Pre-Harvey}                                                                                   & \multicolumn{2}{l|}{Post-Harvey}                                                                                      & \multicolumn{2}{l|}{System   Recovery}                                                                            \\ \cline{2-7} 
                                                                           & \begin{tabular}[c]{@{}l@{}}“Hike”\\ (\%)\end{tabular} & \begin{tabular}[c]{@{}l@{}}Duration\\ (Days)\end{tabular} & \begin{tabular}[c]{@{}l@{}}“Max Drop”\\ (\%)\end{tabular} & \begin{tabular}[c]{@{}l@{}}Duration\\ (Days)\end{tabular} & \begin{tabular}[c]{@{}l@{}}“Peak”\\ (\%)\end{tabular} & \begin{tabular}[c]{@{}l@{}}Duration\\ (Days)\end{tabular} \\ \hline
Gasoline Stations                                                          & 50.82                                                 & 4                                                         & -55.32                                                    & 7                                                         & 25.15                                                 & 34                                                        \\ \hline
Grocery and Merchandise                                                    & 61.35                                                 & 3                                                         & -76.49                                                    & 5                                                         & N/A                                                   & N/A                                                       \\ \hline
\begin{tabular}[c]{@{}l@{}}Health and Personal \\ Care Stores\end{tabular} & 33.06                                                 & 4                                                         & -64.39                                                    & 9                                                         & 16.07                                                 & 30                                                        \\ \hline
Medical Facilities                                                         & 13.15                                                 & 3                                                         & -70.55                                                    & 9                                                         & 23.99                                                 & 28                                                        \\ \hline
Public Order                                                               & N/A                                                   & N/A                                                       & N/A                                                       & N/A                                                       & 72.27                                                 & 45                                                        \\ \hline
Insurance Agencies                                                         & 31.03                                                 & 1                                                         & -63.36                                                    & 7                                                         & 52.02                                                 & 21                                                        \\ \hline
\begin{tabular}[c]{@{}l@{}}Building Material and \\ Supplies\end{tabular}  & 47.99                                                 & 2                                                         & -79.71                                                    & 5                                                         & 39.42                                                 & 58                                                        \\ \hline
Banks                                                                      & 20.54                                                 & 4                                                         & -67.08                                                    & 7                                                         & 33.67                                                 & 26                                                        \\ \hline
Postal Service                                                             & N/A                                                   & N/A                                                       & -90.27                                                    & 6                                                         & 77.54                                                 & 34                                                        \\ \hline
Religious Organizations                                                    & 19.72                                                 & 5                                                         & -73.12                                                    & 5                                                         & 34.57                                                 & 58                                                        \\ \hline
Stores and Dealers                                                         & 17.88                                                 & 1                                                         & -79.57                                                    & 8                                                         & 13.91                                                 & 15                                                        \\ \hline
Restaurants                                                                & 10.94                                                 & 1                                                         & -76.68                                                    & 7                                                         & 17.80                                                 & 34                                                        \\ \hline
Entertainment                                                              & N/A                                                   & N/A                                                       & -77.91                                                    & 10                                                        & N/A                                                   & N/A                                                       \\ \hline
Self-Care                                                                  & 17.73                                                 & 4                                                         & -64.88                                                    & 6                                                         & 37.39                                                 & 22                                                        \\ \hline
\begin{tabular}[c]{@{}l@{}}Recreation and Gym \\ Centers\end{tabular}      & 8.63                                                  & 3                                                         & -79.81                                                    & 7                                                         & 18.12                                                 & 29                                                        \\ \hline
Education                                                                  & N/A                                                   & N/A                                                       & -87.33                                                    & 16                                                        & 40.55                                                 & 24                                                        \\ \hline
\end{tabular}
\end{table}

% \usepackage{multirow}
\begin{table}[]
\renewcommand\thetable{B2} 
   \centering
   \caption{Statistical Information of Resilience Metrics of Different Categories}
\begin{tabular}{|l|l|l|l|l|}
\hline
POI categories                                   & Resilience index      & Mean    & Median & Standard Deviation \\ \hline
\multirow{2}{*}{Gasoline Stations}               & Systematic Impact     & -26.65  & -26.14 & 19.95              \\ \cline{2-5} 
                                                 & Total Recovery Effort & 13.52   & 11.00  & 13.23              \\ \hline
\multirow{2}{*}{Grocery and Merchandise}         & Systematic Impact     & -26.93  & -25.28 & 22.26              \\ \cline{2-5} 
                                                 & Total Recovery Effort & 12.41   & 9.00   & 12.87              \\ \hline
\multirow{2}{*}{Stores and Dealers}              & Systematic Impact     & -37.53  & -38.45 & 20.95              \\ \cline{2-5} 
                                                 & Total Recovery Effort & 17.14   & 15.0   & 13.69              \\ \hline
\multirow{2}{*}{Banks}                           & Systematic Impact     & -30.83  & -28.57 & 30.54              \\ \cline{2-5} 
                                                 & Total Recovery Effort & 11.65   & 9.00   & 13.69              \\ \hline
\multirow{2}{*}{Building Material and Supplies}  & Systematic Impact     & -30.87  & -30.18 & 27.92              \\ \cline{2-5} 
                                                 & Total Recovery Effort & 11.44   & 9.00   & 13.45              \\ \hline
\multirow{2}{*}{Education}                       & Systematic Impact     & -53.97  & -60.12 & 28.62              \\ \cline{2-5} 
                                                 & Total Recovery Effort & 12.48   & 13.00  & 7.15               \\ \hline
\multirow{2}{*}{Entertainment}                   & Systematic Impact     & -28.944 & -16.94 & 31.93              \\ \cline{2-5} 
                                                 & Total Recovery Effort & 10.85   & 6.00   & 14.69              \\ \hline
\multirow{2}{*}{Health and Personal Care Stores} & Systematic Impact     & -38.02  & -39.21 & 25.84              \\ \cline{2-5} 
                                                 & Total Recovery Effort & 14.98   & 13.00  & 11.68              \\ \hline
\multirow{2}{*}{Insurance Agencies}              & Systematic Impact     & -24.6   & 0.00   & 31.48              \\ \cline{2-5} 
                                                 & Total Recovery Effort & 8.95    & 0.00   & 14.47              \\ \hline
\multirow{2}{*}{Medical Facilities}              & Systematic Impact     & -33.66  & -33.39 & 30.54              \\ \cline{2-5} 
                                                 & Total Recovery Effort & 13.31   & 10.0   & 15.52              \\ \hline
\multirow{2}{*}{Recreation and Gym Stores}       & Systematic Impact     & -49.78  & -54.22 & 26.61              \\ \cline{2-5} 
                                                 & Total Recovery Effort & 18.91   & 17.00  & 14.54              \\ \hline
\multirow{2}{*}{Religious Organizations}         & Systematic Impact     & -37.13  & -35.93 & 29.14              \\ \cline{2-5} 
                                                 & Total Recovery Effort & 12.68   & 10.00  & 13.91              \\ \hline
\multirow{2}{*}{Restaurants}                     & Systematic Impact     & -38.93  & -40.99 & 19.88              \\ \cline{2-5} 
                                                 & Total Recovery Effort & 17.40   & 15.00  & 12.13              \\ \hline
\multirow{2}{*}{Self-Care}                       & Systematic Impact     & -32.67  & -31.78 & 29.23              \\ \cline{2-5} 
                                                 & Total Recovery Effort & 13.09   & 10.00  & 14.44              \\ \hline
\end{tabular}
\end{table}

\section{Spatial Distribution of Resilience Metrics across Different POI Categories}
The following figures shows the spatial distribution of the systematic impact and total recovery effort of the emergency preparedness, emergency activity, lifestyle and well-being groups, respectively.

\begin{figure}[ht]
\renewcommand\thefigure{C1} 
\centering
\includegraphics[width=\linewidth]{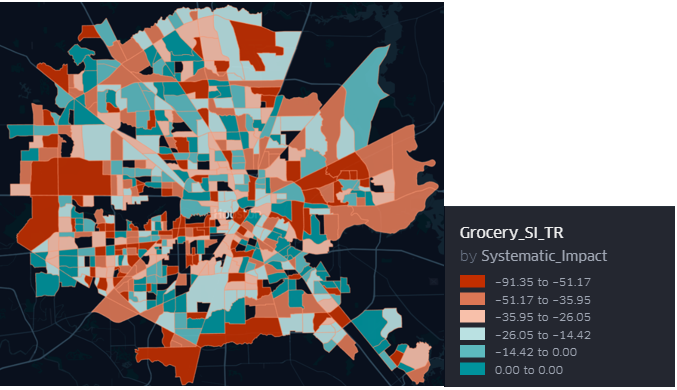}
\caption{Spatial distribution for systemic impact of grocery and merchandise}
\end{figure}

\begin{figure}[ht]
\renewcommand\thefigure{C2} 
\centering
\includegraphics[width=\linewidth]{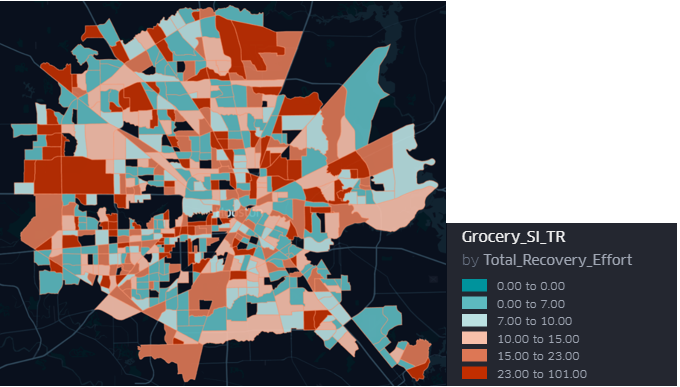}
\caption{Spatial distribution for total recovery effort of grocery and merchandise}
\end{figure}

\begin{figure}[ht]
\renewcommand\thefigure{C3} 
\centering
\includegraphics[width=\linewidth]{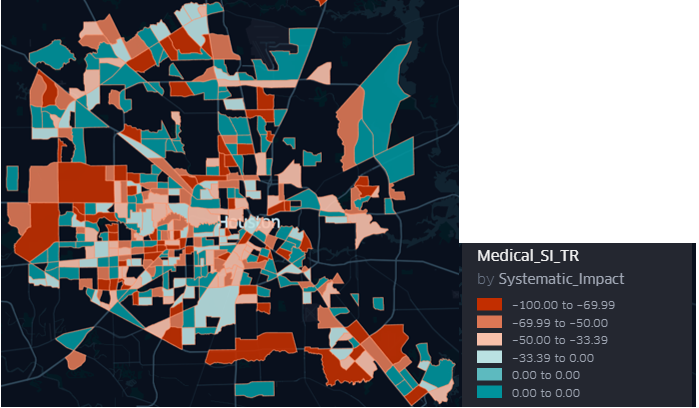}
\caption{Spatial distribution for systematic impact of medical facilities}
\end{figure}

\begin{figure}[ht]
\renewcommand\thefigure{C4} 
\centering
\includegraphics[width=\linewidth]{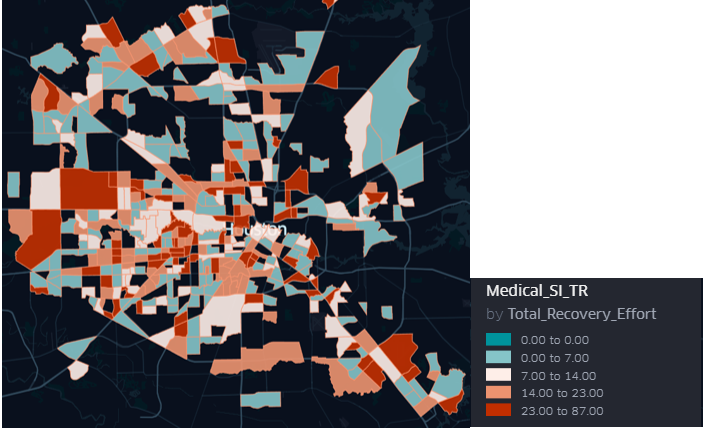}   
\caption{Spatial distribution for total recovery effort of medical facilities}
\end{figure}

\begin{figure}[ht]
\renewcommand\thefigure{C5} 
\centering
\includegraphics[width=\linewidth]{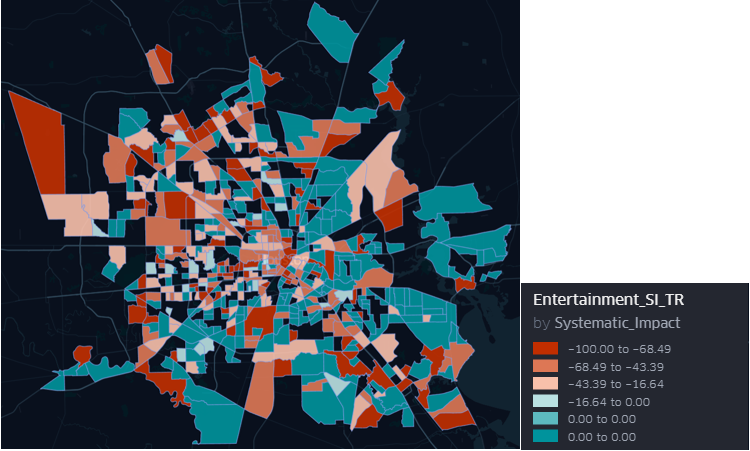}
\caption{Spatial distribution for total recovery effort of entertainment}
\end{figure}

\begin{figure}[ht]
\renewcommand\thefigure{C6} 
\centering
\includegraphics[width=\linewidth]{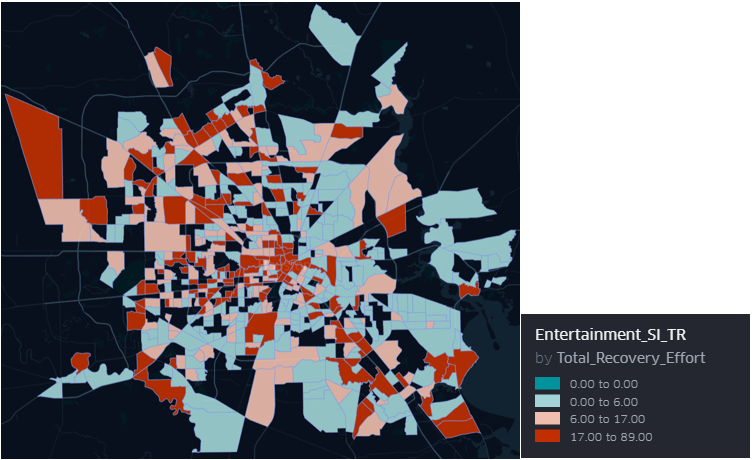}
\caption{Spatial distribution for total recovery effort of entertainment}
\end{figure}

\end{document}